\documentclass[reprint,amsmath,amssymb,aps,floatfix,pra]{revtex4-1}

\usepackage{graphicx}
\usepackage{dcolumn}
\usepackage{bm}
\usepackage{float}
\usepackage{color}

\begin{document}


\title{Effect of magnetic field on the velocity autocorrelation and the caging of particles in two-dimensional Yukawa liquids}

\author{K. N. Dzhumagulova$^1$, R. U. Masheeva$^1$,  T. S. Ramazanov$^1$, Z. Donk\'o$^2$}
\affiliation{$^1$IETP, Al Farabi Kazakh National University, 71, al Farabi av., Almaty, 050040, Kazakhstan}
\affiliation{$^2$Institute for Solid State Physics and Optics,
Wigner Research Centre for Physics,\\
Hungarian Academy of Sciences, H-1121 Budapest, Konkoly-Thege Mikl\'os str. 29-33, Hungary}
\date{\today}

\begin{abstract}
We investigate the effect of an external magnetic field on the velocity autocorrelation function and the ``caging'' of the particles in a two-dimensional strongly coupled Yukawa liquid, via numerical simulations. The influence of the coupling strength on the position of the dominant peak in the frequency spectrum of the velocity autocorrelation function confirms the onset of a joint effect of the magnetic field and strong correlations at high coupling. Our molecular dynamics simulations quantify the decorrelation of the particles' surroundings - the magnetic field is found to increase significantly the caging time, which reaches values well beyond the timescale of plasma oscillations. The observation of the increased caging time is in accordance with findings that the magnetic field decreases diffusion in similar systems. 
\end{abstract}


\maketitle

\section{Introduction}

Strongly coupled plasmas \cite{sccs} comprise a large class of physical systems, in which the ratio of the inter-particle potential energy to the kinetic energy, expressed by the coupling parameter $\Gamma$, (largely) exceeds 1. Dusty plasmas \cite{dusty} are a notable type of strongly coupled many-particle systems that appear both in astrophysical environments and can as well be realized in laboratory. In laboratory settings dust particles can grow in a reactive plasma environment, or can be externally introduced into non-reactive (typically noble gas) discharge plasmas. In this latter case both three-dimensional and two dimensional particle configurations can be realized. Microgravity conditions favor three-dimensional settings, while in the presence of gravity lower-dimensional configurations are routinely formed. 

In typical laboratory setups (a radio-frequency driven plasma source with parallel, horizontal electrodes) two-dimensional layers of particles can be realized \cite{dusty_first}, the position of the dust particle layer is determined by the balance of the major forces acting on the particles, which are usually the electrostatic force, gravitational force, and ion drag force. Additional forces, e.g. the thermophoretic force \cite{NEW} can change the particle configuration drastically, and make it possible to realize three-dimensional structures (Yukawa balls) in the presence of a thermal gradient of the background gas \cite{balls}. A wide variety of physical phenomena taking place in two-dimensional particle layers -- e.g. crystal formation and melting \cite{crystal}, transport processes \cite{transport}, as well as the propagation of waves \cite{waves} --  have been thoroughly investigated in experiments, by theoretical approaches, and via simulation methods. 

Besides the crystallized phase, the strongly coupled liquid phase of dusty plasmas has been receiving a lot of attention. It is an important property of this phase that the surroundings of individual particles are ``quasi-stable'' for a certain time, in contrast to the solid and gaseous limiting cases, where the time of particle localization, respectively, is infinite and extremely short. This property of the liquid phase is the basis of several features of strongly coupled plasmas \cite{Jerome}: among other properties, the Coulomb one-component plasma in the $\Gamma \gtrsim 50$ domain was found (i) to exhibit a  shear viscosity that follows an Arrhenius type behavior, with an activation energy related to the ``binding energy'' of the particles in the cages, and (ii) to obey the Stokes-Einstein relation, characteristics for dense fluids. At strong coupling the particles oscillate in local minima of the potential landscape, which itself, changes on the time scale of particle diffusion. This difference of the time scales for the plasma oscillations and diffusion serves as the basis of the Quasi-Localized Charge Approximation \cite{qlca}, that allows calculation of the dispersion relations of collective excitations from static properties of the system (pair correlation). 

Quantitative data for the localization time have been obtained for Coulomb and Yukawa liquids in \cite{caging} by Molecular Dynamics simulations, using a technique of \cite{Rabani}, that allows tracing of the changes of the neighborhoods of the particles. 
The simulation results have confirmed that at high coupling the particles spend several oscillation cycles in local minima of the potential surface without experiencing substantial changes in their surroundings. 	

The effect of magnetic fields on strongly coupled dusty plasmas became an important topic in the last few years \cite{harmonic_prl,mag_waves,mag_waves2,mag_waves3, mag_diff}. Theoretical and simulation studies have demonstrated the formation of magnetoplasmons and their higher harmonics in strongly coupled Coulomb and Yukawa systems \cite{harmonic_prl}. Detailed studies of the impact of the magnetic field on the collective excitations and the self-diffusion have been presented, respectively, in \cite{mag_waves} and \cite{mag_diff}. The effect of magnetic field on binary Yukawa systems has been studied in \cite{Ott14}. Another line of research focuses on systems of superparamagnetic particles \cite{superpara}.

Experiments, aimed at the realization of magnetized dusty plasmas, have faced, however, serious difficulties, as the external magnetic fields cause a significant perturbation to the plasma itself (like filamentation) before affecting the dynamics of the dust system \cite{extmag}. An alternative method to investigate magnetic field effects was suggested in \cite{newmag}, based on the equivalence of the magnetic Lorentz force and the Coriolis inertial force acting on moving objects when they are viewed in a rotating reference frame. Experimental realization of a rotating dusty plasma has confirmed the theoretical predictions and has proven the formation of magnetoplasmons \cite{RotoDust} in the ``magnetized'' dusty plasma. We note that in magnetized strongly coupled plasmas many of the effects are qualitatively different from those observed and well known for weakly coupled plasmas, due to the interplay of magnetization and strong correlation effects. 

In this paper we investigate the effect of an external, homogeneous magnetic field on the behavior of the velocity autocorrelation function (VACF) in the time and frequency domains, as well as on the caging of the particles, in two-dimensional strongly coupled Yukawa liquids. These phenomena are investigated using Molecular Dynamics simulations. The model and the simulation techniques are described in Sec. II. In Sec. III we present and analyze the simulation results, while Sec. IV gives a short summary of the work.

\section{Model and simulation method}

We investigate the effect of the magnetic field on many-particle systems, in which particles interact via a screened Coulomb (Debye-H\"uckel, or Yukawa) potential:
\begin{equation}
\phi(r) = \frac{Q}{4 \pi \varepsilon_0} \frac{\exp(-r/\lambda_D)}{r},
\end{equation}
where $Q$ is the charge of the particles and $\lambda_D$ is the screening (Debye) length. The ratio of the inter-particle potential energy to the thermal energy is expressed by the coupling parameter
\begin{equation}
\Gamma = \frac{Q^2}{4 \pi \varepsilon_0 a k_B T},
\end{equation}
where $T$ is temperature. We introduce the screening parameter $\kappa = a / \lambda_D$, where $a = (1/\pi n)^{-1/2}$ is the  two-dimensional Wigner-Seitz radius and $n$ is the areal number density of the particles. 

In particular, we investigate the effect of the magnetic field on the velocity autocorrelation function (VACF) of the particles and the cage correlation function that quantifies the relation of the localization time of the particles to the timescale of plasma oscillations. 

We apply the Molecular Dynamics (MD) simulation method to describe the motion of the particles governed by the Newtonian equation of motion. For the integration of the equation of motion that accounts for the presence of the magnetic field we use the method described in \cite{integration}. The number of particles is fixed at $N$ = 4000 (at $N$ = 1000 in the calculations of cage correlations, see later) and we use a quadratic simulation box. The particles move in the $(x,y)$ plane and the magnetic field is assumed to be homogeneous and directed perpendicularly to the two-dimensional layer of the particles, i.e. ${\bf B} = (0,0,B)$. The strength of the magnetic field is expressed in terms of 
\begin{equation}
\beta = \frac{\omega_c}{\omega_p},
\end{equation}
where $\omega_c =  Q B / m$ is the cyclotron frequency and $\omega_p= \sqrt{n Q^2 / 2 \varepsilon m a}$ is the nominal 2D plasma frequency. We note that the Larmor radius becomes smaller than the WS radius at magnetic fields $\beta \gtrsim 0.1.$

The velocity autocorrelation function is defined as (see e.g. \cite{Hamaguchi}):
\begin{equation}
A_{vv}(t) = \langle {\bf v}(t) {\bf v}(0) \rangle,
\end{equation}
while its normalized value (giving $A_{vv}(0) =1$) is given by:
\begin{equation}
\overline{A}_{vv}(t) = \frac {\langle {\bf v}(t) {\bf v}(0) \rangle } {\langle {\bf v}(0) {\bf v}(0) \rangle }.
\end{equation}
The Fourier transform of the VACF is defined as
\begin{equation}
A_{vv}(\omega) = \int_0^\infty A_{vv}(t) {\rm e}^{i\omega t} {\rm d}t,
\label{eq:ft}
\end{equation}
and is calculated by replacing the upper limit of the integration with a time $t_{max}$, for which $A_{vv}(t)\cong 0$ at $t > t_{max}$. 

Taking the time integral of the VACF the self-diffusion coefficient of the particles can be calculated: 
\begin{equation}
D = \frac{1}{2} \int_0^\infty A_{vv}(t) {\rm d}t.
\label{eq:diff}
\end{equation}
We note, however, that previous studies have shown that this integral may be divergent for 2D system, at certain range of parameters \cite{corr}, where calculations, as well as experimental measurements of the self diffusion coefficient, based on the mean square displacement (MSD) of the particles have both shown superdiffusion, MSD$\propto t^\alpha$, with $\alpha > 1$ \cite{superdiffusion}. 

The Fourier transform of the VACF is known to be connetcted with the longitudinal and transverse fluctuations in the system \cite{Hamaguchi}. Therefore we also calculate the respective fluctuation spectra $L(k,\omega)$ and $T(k,\omega)$, for a discrete set of wave numbers $k = m (2 \pi / L) = m k_{min},~m=1,2,...$, accommodated by the simulation box of edge length $L$. To accomplish this calculation we collect data during each time step of the simulation for the microscopic currents 
\begin{eqnarray}
\lambda(k,t)= \sum_j v_{j x}(t) \exp \bigl[ i k x_j(t) \bigr], \nonumber \\
\tau(k,t)= \sum_j v_{j y}(t) \exp \bigl[ i k x_j(t) \bigr],
\label{eq:dyn}
\end{eqnarray}
where $x_j$ and $v_j$ are the position and velocity of the $j$-th particle. These data sequences are subsequently Fourier analyzed to yield, e.g. $L(k,\omega)$ as:
\begin{equation}\label{eq:sp1}
L(k,\omega) = \frac{1}{2 \pi N} \lim_{\Delta T \rightarrow \infty}
\frac{1}{\Delta T} | \lambda(k,\omega) |^2,
\end{equation}
where $\Delta T$ is the length of data recording period and $\lambda(k,\omega) = {\cal{F}} \bigl[ \lambda(k,t) \bigr]$ is the Fourier transform of (\ref{eq:dyn}). Calculation of $T(k,\omega)$ proceeds in the same way. We note that the collective modes show up as peaks in these current fluctuation spectra. 

To quantify the time-dependence of the correlation of the particles' surroundings we adopt the method proposed in \cite{Rabani} and used in \cite{caging} for the investigation of strongly coupled Coulomb and Yukawa liquids. We use a generalized neighbor list $\ell_i$ for particle $i$, $\ell_i = \{f_{i,1},f_{i,2},...,f_{i,N} \}$. Due to the sixfold symmetry we always find the six closest neighbors of particle $i$ and the $f$-s corresponding to these particles are set to a value 1, while all other $f$-s are set to 0.

The similarity between the surroundings of the particles at $t$=0 and $t>0$ is measured by the {\it list correlation function}:
\begin{equation}
C_{\ell}(t) = {{\langle \ell_i(t) \ell_i(0) \rangle} \over
{\langle \ell_i(0)^2 \rangle}},
\label{eq:listcorr}
\end{equation}
where $\langle\cdot\rangle$ denotes averaging over particles and initial times. $C_{\ell}(t=0) = 1$, and $C_{\ell}(t)$ is a monotonically decaying function (provided that averaging is sufficient).

The number of particles that have left the original cage of particle $i$ at time $t$ can be determined as
\begin{equation}
n_i^{\rm out} (t) = |\ell_i(0)^2| - \ell_i(0) \ell_i(t),
\label{eq:n}
\end{equation}
where the first term gives the number of particles around particle $i$ at $t$ = 0 (that, actually, equals to six, in our case), while the second term gives the number of `original' particles that remained in the surrounding after time $t$ elapsed. As the next step an integer value $c$ is defined, which is the number of the 'original' neighbors that have to leave the cage before we say that the cage has undergone a 'substantial change'. The {\it cage correlation function} $C_{\rm cage}^{c}(t)$ can be calculated by averaging over particles and initial times, of the function
$\Theta (c - n_i^{\rm out})$, i.e.
\begin{equation}
C_{\rm cage}^{c}(t) =\langle\Theta(c - n_i^{\rm out} (0,t)) \rangle.
\label{eq:cagecorr}
\end{equation}
Here $\Theta$ is the Heaviside function. We calculate the cage correlation functions for $c=3$, meaning that half of the original neighbors leave the cage. We adopt the definition of the {\it caging time} introduced in \cite{caging}, according to which $t_{\rm cage}$ is defined as the time when $C_{\rm cage}^{3}$ decays to a value 0.1.

\section{Results}

\subsection{Velocity autocorrelation}

\begin{figure}[h!]
\includegraphics[width=0.85\columnwidth]{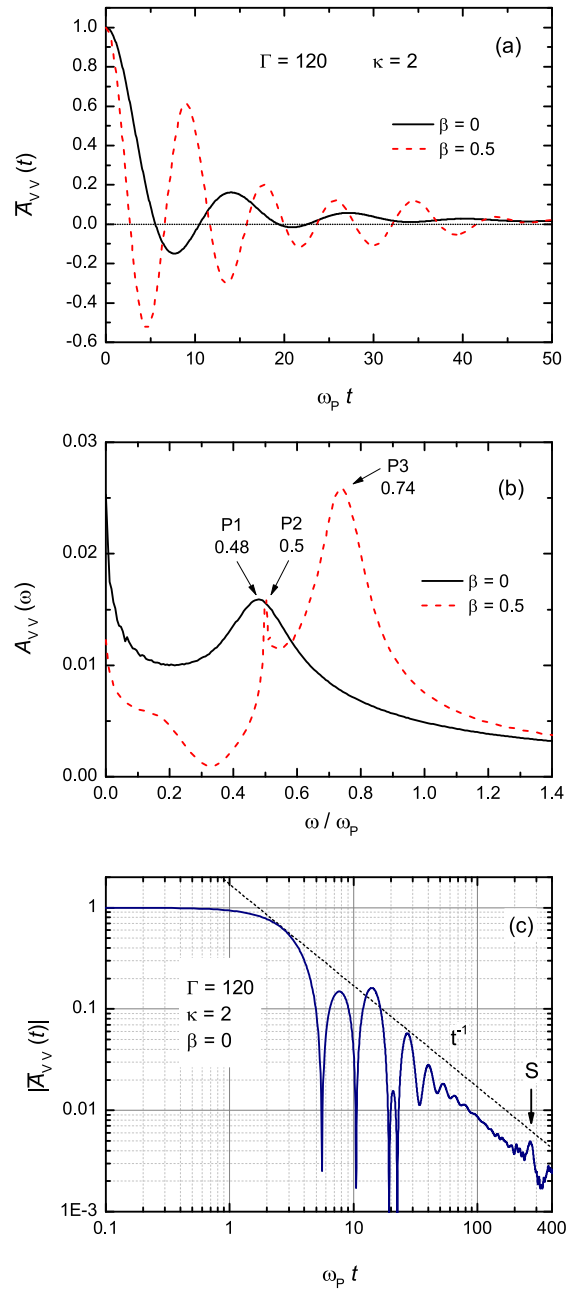}
\caption{\label{fig:vac1}
(color online) (a) Normalized velocity autocorrelation functions at $\beta=0$ and $\beta=0.5$, for $\Gamma=120$ and $\kappa=2$. Fourier transform of the VACF-s shown in (a). P1, P2, and P3 identify peaks of the spectra, the peak positions are given in units of $\omega_p$. (c) Magnitude of the normalized VACF for the above $\Gamma$ and $\kappa$ values, for the unmagnetized case. Note the closely $t^{-1}$ decay of the curve at long times. S denotes the ``sound peak'' originating from the finite size of the simulation cell, see \cite{corr}.}
\end{figure}

\begin{figure}[h!]
\includegraphics[width=0.85\columnwidth]{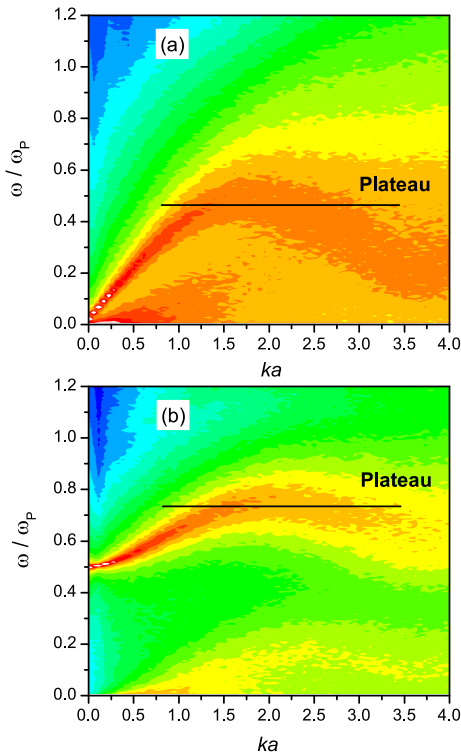}
\caption{\label{fig:vac2}
(color online) The sum of longitudinal and transverse current fluctuation spectra, $L(k,\omega)+T(k,\omega)$, at $\Gamma=120$ and $\kappa=2$, for (a) the unmagnetized and (b) magnetized, $\beta=0.5$ system. Note the lifting of the longitudinal mode in the magnetized system. The well-recognizable modes correspond to the longitudinal current fluctuations, the transverse current fluctuations appear with weak amplitude, at low frequencies (at $\omega / \omega_p \lesssim 0.2$).}
\end{figure}

\begin{figure}[htb]
\includegraphics[width=0.85\columnwidth]{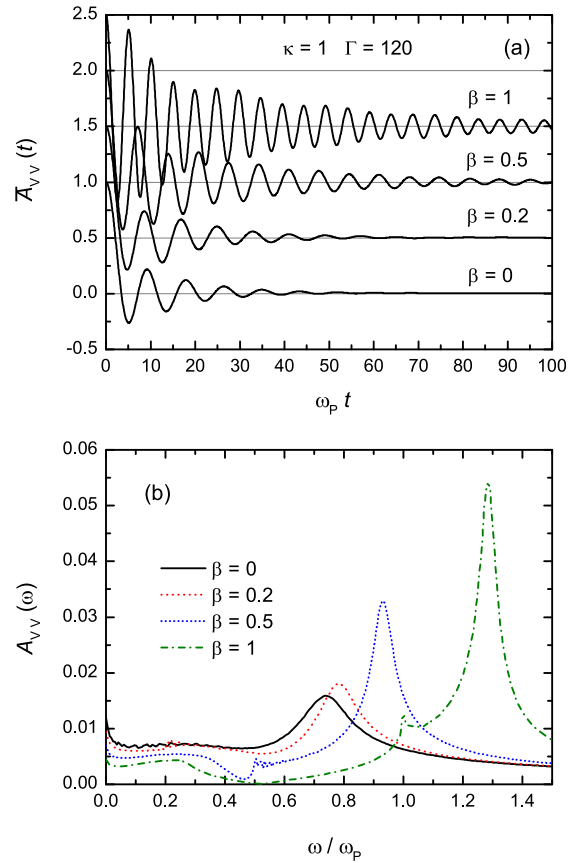}
\caption{\label{fig:vac3}
(color online) Normalized VACF-s for a set of magnetization values at $\Gamma=120$ and $\kappa=1$, and their frequency domain behavior, $A_{vv}(\omega)$. Note that the curves in (a) are vertically shifted, for the clarity of the plot.}
\end{figure}

\begin{figure}[htb]
\includegraphics[width=0.85\columnwidth]{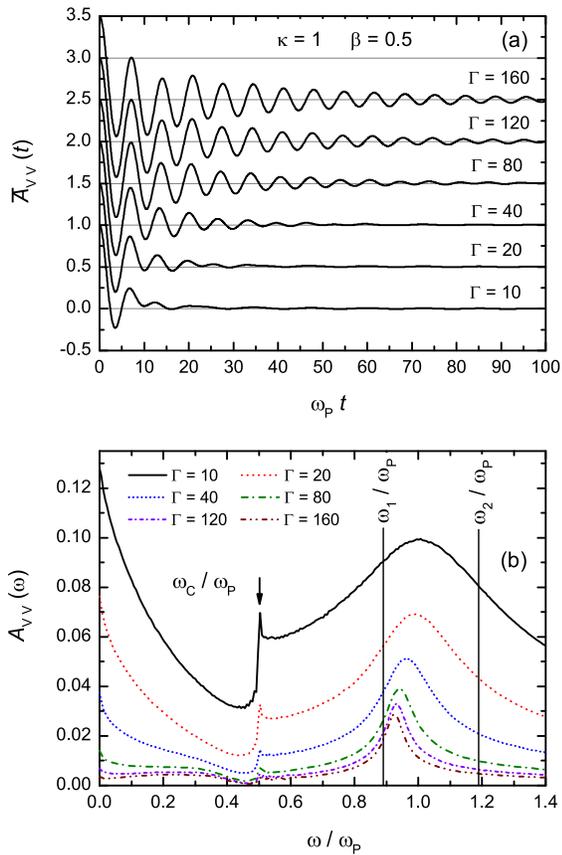}
\caption{\label{fig:vac4}
(color online) Normalized VACF-s for a series of $\Gamma$ values at fixed values of $\kappa=1$ and $\beta=0.5$, and (b) their frequency domain behavior, $A_{vv}(\omega)$. Note that the curves in (a) are vertically shifted, for the clarity of the plot. The vertical solid lines in panel (b) indicate the frequencies given by Eqs. (\ref{eq:omega1}) and (\ref{eq:rpa}), respectively, for the strong and weak coupling limits.}
\end{figure}

\begin{figure}[htb]
\includegraphics[width=0.85\columnwidth]{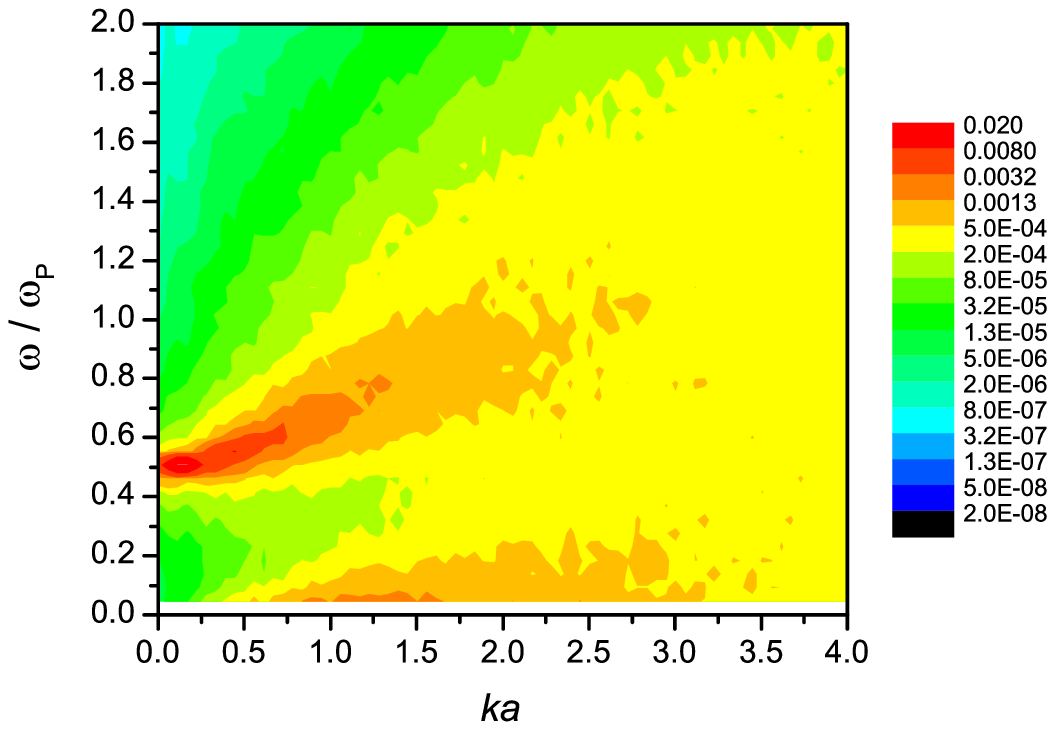}
\caption{\label{fig:spectrum2}
(color online) The sum of longitudinal and transverse current fluctuation spectra, $L(k,\omega)+T(k,\omega)$, at $\Gamma=10$, $\kappa=2$, and $\beta=0.5$ system.}
\end{figure}

The general effect of the external magnetic field on the velocity autocorrelation function is illustrated in Fig.~\ref{fig:vac1}(a), at $\Gamma=120$ and $\kappa=2$. At $\beta=0$ the VACF exhibits a few oscillations, which is a fingerprint of localized oscillations of the particles. In the magnetized case the dominant frequency is clearly increased, as well as the values of the extrema of the oscillatory VACF. The Fourier transform of the VACF, $A_{vv}(\omega)$, shown in Fig.~\ref{fig:vac1}(b) exhibits characteristic changes when the magnetic field is applied. In the $\beta=0$ case the spectrum exhibits a single peak at about $\omega / \omega_p \approx 0.48$. This peak is known the be related to the longitudinal current fluctuations (see e.g. \cite{Hamaguchi}), and this frequency indeed corresponds to the plateau of the dispersion relation of the longitudinal mode, shown in Fig.~\ref{fig:vac2}(a). At $\beta>0$ the formation of a magnetoplasmon shifts the peak position to a higher value, $\omega / \omega_p \approx 0.74$, following the change of the character of the mode dispersion curve, plotted in Fig.~\ref{fig:vac2}(b). The low frequency part of $A_{vv}(\omega)$, seen in Fig.~\ref{fig:vac1}(b), is depleted at $\beta>0$ with respect to the $\beta=0$ case, and we also observe the formation of a small peak at $\omega / \omega_p = 0.5$, corresponding to the cyclotron frequency of the dust particles. 

We note that, according to (\ref{eq:ft}) and (\ref{eq:diff}), $D = \frac{1}{2} A_{vv}(\omega=0)$. Evaluation of $A_{vv}(\omega=0)$ from Fig.~\ref{fig:vac1}(b) is ambigous, in accordance with the possible divergence of the Green-Kubo integral (\ref{eq:diff}) already quoted. Plotting the normalized VACF for the unmagnetized case (see Fig.~\ref{fig:vac1}(c)) confirms that $\overline{A}_{vv}(t)$ decays as $t^{-1}$ at long times, making (\ref{eq:diff}) divergent (if we assume that the decay rate is maintained up to infinitely long time). It is not the topic of the present paper to investigate this effects further, and, accordingly, we shall not discuss the $\omega \rightarrow 0$ behavior of $A_{vv}(\omega)$.

Fig.~\ref{fig:vac3}(a) presents a series of normalized VACF-s for increasing magnetization, at fixed $\Gamma=120$ and $\kappa=1$. The data show that the behavior of $A_{vv}(t)$ is significantly altered with the introduction of the magnetic field. The dominant frequency (easily observed by eye) increases with increasing $\beta$, and the oscillations of the VACF-s persist for an increasingly longer time when the strength of the magnetic field is increased. Fig.~\ref{fig:vac3}(b) shows the respective $A_{vv}(\omega)$ functions in the frequency domain, where the dominant frequency shows up as a definite peak. In \cite{harmonic_prl} it has been shown that at high coupling the dominant frequency in the longitudinal fluctuation spectrum takes a value 
\begin{equation}
\omega_1^2 =  \omega_c^2 + 2 \omega_E^2 = \beta^2 \omega_p^2 + 2 \omega_E^2,
\label{eq:omega1}
\end{equation}
where $\omega_E$ is the Einstein frequency, defined as the oscillation frequency of a test particle in a frozen environment. At $\kappa=1$ we have $\omega_E \cong 0.52 \omega_p$ (see Fig. 19(b) of Ref. \cite{review}, note, however, that the numerical data of the same paper, given by eq. (54) are false). The positions of the peaks observed in Fig.~\ref{fig:vac3}(b), in comparison with the theoretical prediction given above, are listed in Table 1. We find a very good agreement (only a few \% deviation) between the two sets of data, confirming the theoretical arguments of \cite{harmonic_prl}, according to which the dominant oscillation frequency forms due to a combined effect of magnetic field and strong correlations, lifting the fundamental frequency above the cyclotron frequency $\omega_c = \beta \omega_p$.

Next, we investigate the effect of the coupling strength on the normalized VACF-s and their Fourier transform, at a fixed value of normalized magnetic field, $\beta=0.5$. The data are displayed in Fig.~\ref{fig:vac4}, for $\kappa=1$. The time-domain data show only a more persisting correlation at higher coupling, however, the frequency spectrum $A_{vv}(\omega)$ shows an upwards shift of the dominant peak with lowering $\Gamma$. The relation (\ref{eq:omega1}), discussed above, holds only for a high coupling. In the limit of vanishing correlations (weakly coupled plasma limit, $\Gamma \rightarrow 0$) the frequency of the resulting (upper) hybrid mode in a magnetized plasma (where the direction of propagation is perpendicular to the direction of the magnetic field) is known to turn into the Random Phase Approximation (RPA) value (for details see \cite{Hou}),
\begin{equation}
\omega_2^2 = \omega_c^2 + \omega_p^2 = \omega_p^2 (\beta^2+1).
\label{eq:rpa}
\end{equation}
At $\beta=0.5$, the frequency defined by the above equation is $\omega_2 \cong 1.19 \omega_p$. This frequency value is not reached by our simulation data at decreasing coupling, due to the significant broadening of the frequency spectrum, as indicated in Fig.~\ref{fig:spectrum2}, for the conditions $\Gamma=10$, $\kappa=2$, and $\beta=0.5$. Here the magnetoplasmon becomes hardly recognizable at reduced wave numbers $ka \gtrsim 2.5$, where the fluctuation spectrum is practically featureless.

\begin{table}[htdp]
\caption{The dependence of the frequency of the dominant peak observed in $A_{vv}(\omega)$, as function of the normalized magnetic field $\beta$, for $\Gamma=120$ and $\kappa=1$.}
\begin{center}
\begin{tabular}{|c|c|c|}
\hline
$\beta$ & Observed $\omega/\omega_p$ & Theoretical $\omega_1/ \omega_p $, Eq. (\ref{eq:omega1})\\
\hline
0 & 0.74 & 0.74 \\
0.2 & 0.78 & 0.76 \\
0.5 & 0.92 & 0.89 \\
1.0 & 1.28 & 1.24 \\
\hline
\end{tabular}
\end{center}
\label{tab:1}
\end{table}

\begin{figure}[htb]
\includegraphics[width=0.85\columnwidth]{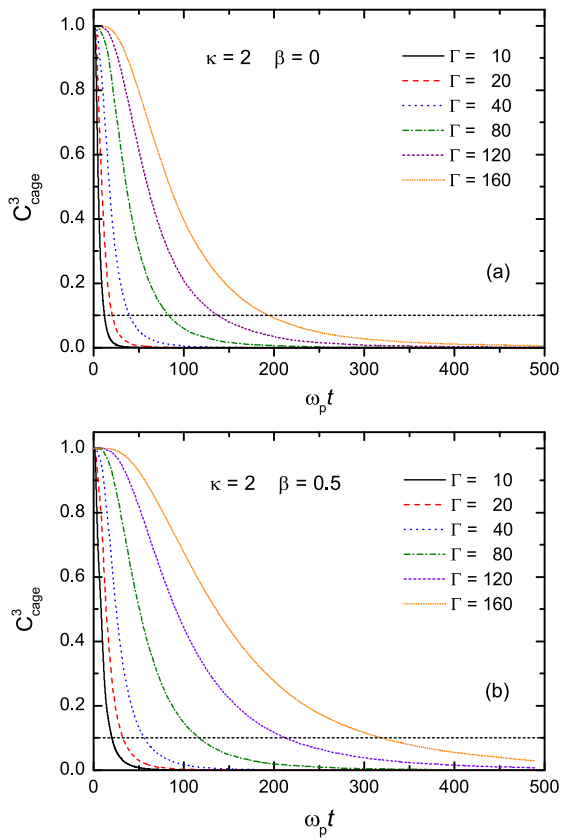}
\caption{\label{fig:cage1}
(color online) Cage correlation functions: the effect of $\Gamma$ at (a) $\beta=0$ and (b) $\beta=0.5$. }
\end{figure}

\begin{figure}[htb]
\includegraphics[width=0.85\columnwidth]{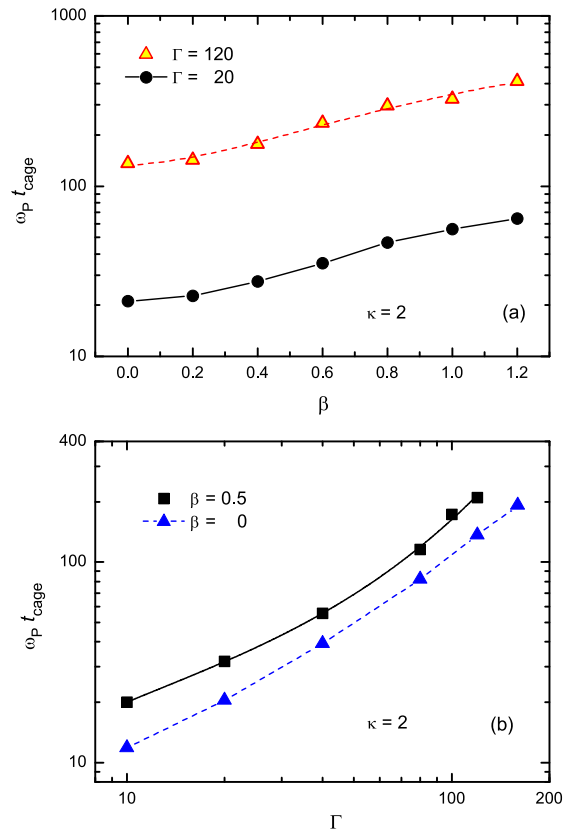}
\caption{\label{fig:time}
(color online) The dependence of the caging time (a) on the magnetic field strength at $\Gamma=20$ and 120, and (b) on the coupling strength $\Gamma$ at $\beta=0$ and 0.5.}
\end{figure}

\subsection{Caging}

The dependence of the cage correlation function $C_{\rm cage}^3(t)$, defined by eq. (\ref{eq:cagecorr}), on the system parameters, is analyzed in Fig.~\ref{fig:cage1}. The cage correlation functions have been calculated for a series of $\Gamma$ values, for the unmagnetized case ($\beta=0$) and for a moderate value of the magnetization, $\beta=0.5$. The data are shown for $\kappa=2$, the behavior is similar at other values of screening. Comparison of the data for these two cases, shown in Fig.~\ref{fig:cage1}(a) and (b), respectively,  reveals the increase of the caging time with increasing magnetic field. This behavior can easily be understood by the decreasing Larmor radius of the particles, that becomes a fraction of the interparticle distance at the highest $\beta$ values considered.

Finally, we show the dependence of the caging time, defined earlier as $C_{\rm cage}^3(t_{\rm cage}) =0.1$, in Fig.~\ref{fig:time}(a) on the magnetic field (at $\kappa=2$ and coupling values $\Gamma=120$ and 20), and in Fig.~\ref{fig:time}(b) on $\Gamma$ (at fixed $\kappa=2$ and magnetic field strengths $\beta=0$ and 0.5). Fig.~\ref{fig:time}(a) reveals an approximately three times increase of the caging time when $\beta$ is increased from 0 to 1, at both values of coupling. 

At the parameter pair $\Gamma=120$ and $\kappa=2$ we reach $\omega_p t_{\rm cage} \cong 400$ at $\beta=1.2$. Now we estimate how many oscillations caged particles execute during this time. One oscillation cycle of a caged particle in the strong coupling domain corresponds to $\omega_1 t \cong 2 \pi$, where $\omega_1$ is defined by Eq.~(\ref{eq:omega1}). For a number of oscillation cycles, $N_{osc}$, within the cage $\omega_1 t_{\rm cage} \cong 2 \pi N_{osc}$ holds. from this,
\begin{equation}
N_{osc} = \frac{1}{2\pi} (\omega_p t_{\rm cage}) \frac{\omega_1}{\omega_p} = 
\frac{1}{2\pi} (\omega_p t_{\rm cage}) \sqrt{ \beta^2 + 2\frac{\omega_E^2}{\omega_p^2}}. 
\end{equation}
For $\kappa=2$ we have $\omega_E / \omega_p \cong 0.32$ (according to Ref. \cite{review}), so at $\beta=0$ and $\beta=0.5$, respectively, $N_{osc} \cong 0.051 (\omega_p t_{\rm cage})$, and  $N_{osc} \cong 0.107 (\omega_p t_{\rm cage})$. Selected values, corresponding to the data shown in Fig.~\ref{fig:time}(b), are given in Table II.

The data unambiguously confirm that in the $\Gamma \gg 1$ domain the particles carry out several oscillation cycles within their cages, before the potential landscape changes due to the diffusion of the particles. In magnetized systems $N_{osc}$ increases because of two reasons: (i) due to the reduction of the diffusion with increasing magnetic field, and (ii) due to the increasing oscillation frequency in the strong coupling domain. This effect has important consequences in determining the properties of strongly coupled plasmas \cite{Jerome}.

\begin{table}[h!]
\caption{Number of oscillations cycles of caged particles, $N_{osc}$, at selected $\Gamma$, $\beta$ parameter pairs, at $\kappa=2$.}
\begin{center}
\begin{tabular}{|c|c|c|}
\hline
$\Gamma$ & $\beta=0$ & $\beta=0.5$ \\
\hline
10 & 0.6 & 2.2 \\
40 & 2.0 & 5.9 \\
120 & 6.9 & 22.5 \\
\hline
\end{tabular}
\end{center}
\label{tab:2}
\end{table}

\newpage

\section{Summary}

We have investigated the effect of a homogeneous magnetic field on the velocity autocorrelation function and the caging phenomenon in two-dimensional Yukawa liquids in the strong coupling domain. The velocity autocorrelation functions have been analyzed both in the time and frequency domains. The dominant peak in $A_{vv}(\omega)$, related to the longitudinal current fluctuations in the liquid, was shown to be formed at high coupling at a frequency defined by a joint effect of cyclotron motion and strong inter-particle correlations \cite{harmonic_prl}. Towards lower coupling values the position of the peak was found to shift upwards, towards the RPA limit, which, however, was not reached due to the broadening of the peak. The caging time of the particles, of which the relation to the plasma oscillation cycles is of paramount importance in determining the liquid state properties of the plasma, was found to increase significantly with the applied magnetic field. Experimental verification of our computational results should be possible in future rotating dusty plasma experiments \cite{RotoDust}.

\begin{acknowledgments}
This work has been supported by the Grants OTKA K-105476 and MES RK 1137/GF-1. We thank Prof. M. Bonitz for useful discussions.
\end{acknowledgments}

\end{document}